\documentstyle[prd,aps,epsf]{revtex}

\newcommand{\bea}{\begin{eqnarray}}
\newcommand{\eea}{\end{eqnarray}}
\newcommand{\beq}{\begin{equation}}
\newcommand{\eeq}{\end{equation}}
\newcommand{\del}{\partial}

\begin{document}

\draft

\twocolumn[\hsize\textwidth\columnwidth\hsize\csname
@twocolumnfalse\endcsname

\title{Evolution of Fluctuations during Graceful Exit in String 
Cosmology}

\author{Shinsuke Kawai${}^{1}$ and Jiro Soda${}^{2}$}
\address{
  ${}^{1}$
  Graduate School of Human and Environmental Studies, Kyoto University,
  Kyoto 606-8501, Japan\\
  {\tt kawai@phys.h.kyoto-u.ac.jp}\\
  ${}^{2}$
  Department of Fundamental Sciences, FIHS, Kyoto University, 
  Kyoto 606-8501, Japan\\
  {\tt jiro@phys.h.kyoto-u.ac.jp} }
\date{\today}
\maketitle

\begin{abstract}
We study the evolution of fluctuations in a universe dominated by a scalar 
field coupled to the Gauss-Bonnet term.
During the graceful exit, we found non-negligible enhancements of both 
curvature perturbation and gravitational wave in the long wavelength limit, 
and we also found a short wavelength instability for steep background 
superinflation just after the completion of the graceful exit.
This result for one possible graceful exit mechanism would provide a 
significant implication on the primordial spectrum from the string cosmology. 

\end{abstract}

\noindent
%
\vskip2pc]
%
{\em 1. Introduction}\\
Recent developments of non-perturbative string theory brought a new viewpoint
in cosmology. 
Pre-big bang scenario based on the scale factor duality was invented to predict
the universe before the big bang. 
This scenario is expected to be confirmed or excluded by the observations, 
where spectrum generated in the very early universe serves as a key test.
In the simple dilaton-metric pre-big bang model\cite{pbb}, the superinflation
driven by the kinetic energy of the dilaton is known to generate a 
spectrum with $n\simeq 4$\cite{blue}, 
which differs significantly from the observationally supported $n\simeq 1$ 
spectrum. 
It is argued that other field contents, e.g. pseudo-scalar axion, 
is required to be included\cite{axion}. 
One of the most exciting possibility pointed out in string cosmology is that 
the gravitational waves from the stringy era might be detectable with the next 
generation gravitational wave interferometers, such as the Laser 
Interferometric Gravitational Observatory (LIGO). 
The dilaton-driven inflation of string cosmology allows the existence of the 
stochastic background gravitational wave which passes the constraint from the 
Cosmic Background Explorer (COBE), and at the same time in the detectable 
region of the next generation interferometers\cite{gwspec,obs}. 
If actually detected, this is supposed to be a strong evidence that the very 
early universe is described by the string theory. 

All of these discussions, however, resort to one assumption:
the constancy of the long wavelength perturbations during the graceful exit. 
Since the post-big bang cosmological solution corresponding to our universe is 
separated by a curvature singularity from the pre-big bang solution,
it is not clear whether the fluctuations generated in the pre-big bang scenario
is viable for the structure formation. 

This notorious graceful exit problem of the string cosmology\cite{gexit} 
is expected to be solved by non-perturbative quantum effects. 
Although the concrete mechanism is currently not known, there are several 
examples of the graceful exit accomplished by the quantum effects.
One of them is a treatment using the Wheeler-de Witt equation, and it is
shown that the transition amplitude from the pre-big bang universe to the 
Friedmann universe is not zero\cite{wdw}.
There is another example by Rey\cite{sjrey}, which is a two-dimensional
dilaton-gravity model with a quantum back reaction. In this model the quantum
effect is calculable through the conformal anomaly. 
There are several four dimensional cosmological models with higher curvature
terms, which realize a non-singular transition\cite{r2cosmology,alphaprime}. 
Among them
the model proposed by Antoniadis, et. al.\cite{art94} is considered as
a four dimensional counterpart of Rey's model. This is based on
the one-loop effective action of orbifold-compactified heterotic 
string\cite{oneloop},
and consists of dilaton and modulus fields besides gravity.
The solution in Einstein frame has an accelerated expansion phase driven by 
the kinetic energy of the modulus. This solution makes a smooth transition to a
Friedmann-like expansion by virtue of the modulus field coupled 
to the Gauss-Bonnet curvature invariant. 
This is not literally a pre-big bang model since the modulus rather
than the dilaton drives the accelerated expansion, and since this acceleration
is realized in Einstein frame, not in dilaton frame. 
However, this model realizes a similar situation to the pre-big bang in that
the universe begins from a pole-like inflation and ends in the Friedmann
universe. Naively one can expect that the transition phase does not affect the
superhorizon scale fluctuation, as is assumed in the pre-big bang scenario.

The purpose of this letter is to study the evolution of the perturbations 
in the model of Antoniadis et.al., and see whether the constancy of the long 
wavelength perturbation really holds during the graceful exit. We hope the
outcome of our study will have implication to the study of pre-big bang
if it has a successful graceful exit mechanism similar to this.


{\em 2. Graceful Exit via Gauss-Bonnet term}\\
The model we treat in this letter is the universe with a scalar field coupled 
to the Gauss-Bonnet invariant, which is essentially the same as the 
aforementioned one-loop model\cite{art94} except that the dilaton field is 
neglected.
It should be noted that in several string models the coupling of the scalar 
field to the gravity we use here is shown to be universal beyond one-loop if 
there exists supersymmetry\cite{r2coupling}.
Thus, assuming the domination of one scalar field coupled to $R_{GB}^2$,
the situation we are presenting here is not limited to one special model.
The action of our model is\cite{rt94}
\beq
{\cal S}=\int dx^4 \sqrt{-g}
\left\{\frac 12 R-\frac 12 \del_\mu\varphi\del^\mu\varphi
  -\frac{\lambda}{16}\xi(\varphi)R_{GB}^2
\right\},
\label{eqn:effaction}
\eeq
where 
$R_{GB}^2=R_{\mu\nu\kappa\lambda}R^{\mu\nu\kappa\lambda}
-4R_{\mu\nu}R^{\mu\nu}+R^2$.
Here,
$\lambda$ is a coupling constant determined by the 4-dimensional trace anomaly
of $N=2$ sectors, and can take both signs according to the relative numbers of
supermultiplets in the theory. It is shown in \cite{art94,rt94,maeda} that
non-singular cosmological solutions appear when $\lambda>0$. Since the absolute
value of $\lambda$ is absorbed in the redefinition of the time scale, 
in what follows we put $\lambda=2$ for simplicity.
For the one-loop string effective action the function $\xi(\varphi)$ 
is an even function of the modulus field $\varphi$ written in terms of  
Dedekind $\eta$ function as
\bea
\xi(\varphi)
&& =-\ln[2e^\varphi\eta^4 (i e^\varphi)]\nonumber\\
&& =-\ln 2 - \varphi + \frac{\pi e^\varphi}{3}-4\sum_{n=1}^{\infty}\ln
(1-e^{-2n\pi e^\varphi}).
\label{eqn:ssxi}
\eea
The existence of the non-singular cosmological solutions is shown by Rizos 
and Tamvakis\cite{rt94} for general $\xi(\varphi)$, as long as $\xi$ grows 
like or faster than $\varphi^2$ as $\varphi\rightarrow\pm\infty$. 
In this letter we mainly concentrate on the coupling of the form 
(\ref{eqn:ssxi}).
From the action (\ref{eqn:effaction}) with the flat Friedmann-Robertson-Walker
metric 
\beq
ds^2=-dt^2+a^2\delta_{ij}dx^idx^j,
\eeq
where $a$ is the scale factor of the universe, one obtains the following 
equations for the background:
\bea
&&\dot\varphi^2=6 H^2 (1-H\dot\xi),
\label{eqn:bgeom1}\\
&&(2\dot H+5H^2)(1-H\dot\xi)+H^2(1-\ddot\xi)=0,\\
&&\ddot\varphi+3H\dot\varphi+3(\dot H+H^2)H^2\xi_{,\varphi}=0.
\eea
$H=\dot a/a$ is the Hubble parameter and the overdot denotes the 
differentiation with respect to the cosmic time $t$.
These background solutions are well studied in \cite{art94,rt94}, and
the numerical analysis of this model for $\xi=\varphi^n$ is found in 
\cite{kanti}.
An example of non-singular background solutions (the Hubble parameter and the 
scalar field) is shown in Fig. 1. The general behavior of the non-singular 
solutions which behave like a Friedmann universe in late time is that it starts
from a Minkowski space, subsequently superinflates as $H\sim(-t)^{-2}$, and 
makes a transition at $\varphi\sim 0$ and then 
goes into a Friedmann-like universe ($H\sim 1/3t$). Since $H$ and 
$\dot\varphi$ keep their signs through the evolution and the system is 
invariant under $\varphi\leftrightarrow-\varphi$, we hereafter set $H>0$ and 
$\dot\varphi>0$. 
The asymptotic behaviors of the modulus field are $\varphi\sim-5\ln|t|$ as 
$t\rightarrow-\infty$ and $\varphi\sim\sqrt{2/3}\ln t$ as
$t\rightarrow\infty$. The scale factor approaches a constant $a\sim a_p$
in the asymptotic past, and grows as $a\sim t^{1/3}$ in the asymptotic
future.
Of course, the asymptotic solutions may differ from this if we consider more
realistic model with other matter contents which dominate the system in the
asymptotic regions.
If $\xi(\varphi)=\varphi^2$, the past asymptotic solution becomes de 
Sitter-like, though the behavior near the transition is similar.
\begin{figure}
\epsfxsize=80mm
\epsfysize=55mm
\epsffile{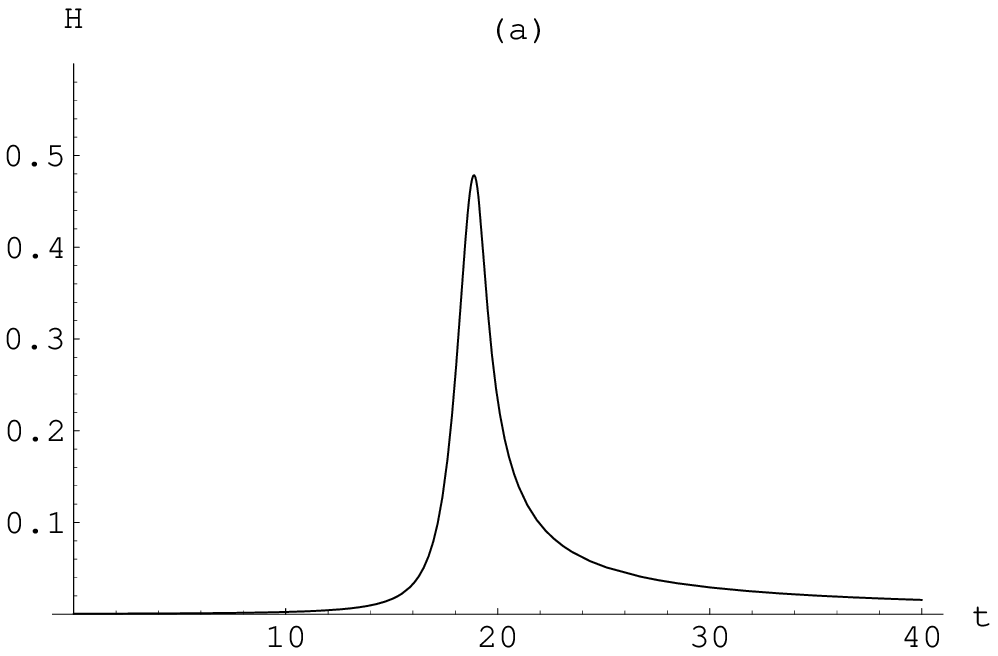}
\vspace{-7mm}
\epsfxsize=80mm
\epsfysize=55mm
\epsffile{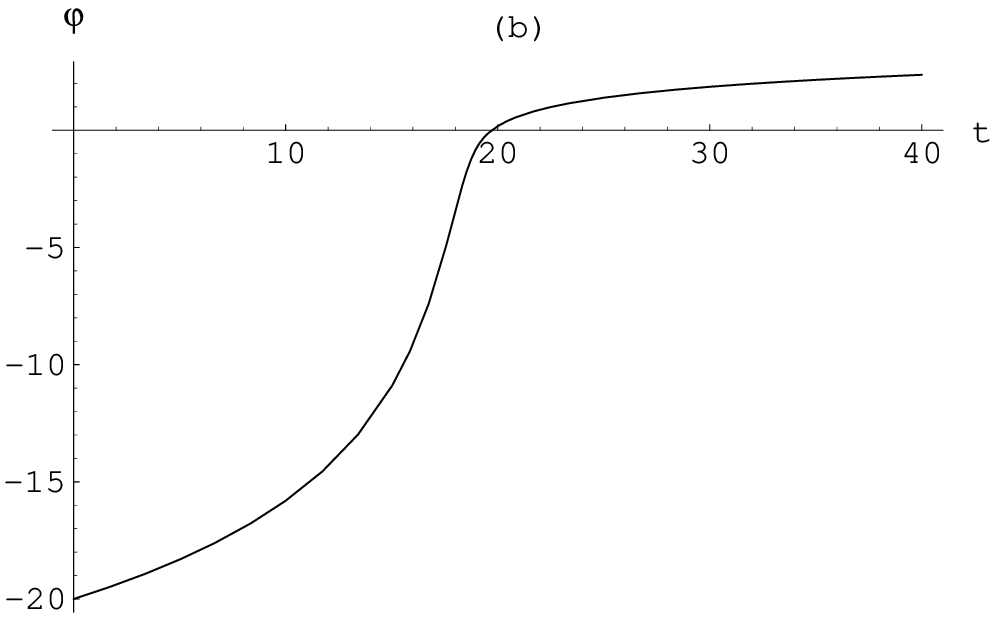}
\vspace{-7mm}
\caption{Habble parameter and scalar field in a non-singular solution. 
The accelerated expansion in the past region is driven by the kinetic
energy of the modulus. The universe makes a transition to Friedmann-like
solution around $\varphi\sim 0$, where the modulus slows down.}
\end{figure}

{\em 3. Scalar perturbation}\\
To calculate the fluctuation we expand the metric and the modulus field
up to linear order of perturbation:
\bea
ds^2&=&
-(1+2\phi)dt^2+2a(V_{|i}-S_{i})dt dx^i\nonumber\\
&&+a^2\left\{(1-2\psi)\gamma_{ij}+2E_{|ij}+2F_{i|j}+h_{ij}
  \right\}dx^idx^j,\\
\label{eqn:metric}
\varphi&=&\varphi^{(0)}+\delta\varphi.
\eea
We will omit the superscript of $\varphi^{(0)}$ in the following.
Due to the symmetry of the background geometry we can decompose the metric 
perturbations into scalar ($\phi$, $\psi$, $V$, $E$), vector ($S_i$, $F_i$) 
and tensor ($h_{ij}$) parts.
All of these perturbations are assumed to be small compared to the 
background of order unity.
The equations for the scalar perturbation becomes particularly simple if we 
use the uniform field gauge ($\delta \varphi=0$, $E=0$):
\bea
&&3H(3\alpha-1)\dot\psi-6H^2(1-\alpha)\phi\nonumber\\
&&\hspace{4mm}-2\alpha\frac{\nabla^2}{a^2}\psi
+H(3\alpha-1)\frac{\nabla^2}{a^2}V=0,\\
&&(3\alpha-1)H\phi+2\alpha\dot\psi=0,\\
&&a\alpha\left\{\phi+(5-3\Gamma)\psi\right\}+(a^2\alpha V)\dot{}=0,
\eea
which are time-time, time-space, traceless space-space part of the Einstein 
equations for perturbations.
$\alpha$ and $\Gamma$ are functions of the background variables, 
defined as $\alpha=1-H\dot\xi$ and $\Gamma=-2\dot H/3 H^2$.
The physical meaning of $\alpha$ is (proportional to) the fraction of the 
modulus kinetic energy to the geometrical kinetic energy, 
which can be seen from (\ref{eqn:bgeom1}). 
$\Gamma$ is the {\em effective adiabatic index} $\Gamma-1=-G^i{}_i/3G^0{}_0$,
where $G^\mu{}_\nu$ is the background Einstein tensor.
They behave as $\alpha\sim t^2$, $\Gamma\sim t$ in the past asymptotic region
and $\alpha\sim 1$, $\Gamma\sim 2$ in the future asymptotic region.
In the uniform field gauge, $\psi$ is the curvature perturbation
in terms of which the equation becomes\cite{hwang}
\beq
\ddot\psi+\frac{\dot A}{A}\dot\psi+B\frac{\nabla^2}{a^2}\psi=0,
\label{eqn:spert}
\eeq
where
\bea
&&A=\frac{a^3\alpha(5\alpha^2-2\alpha+1)}{(3\alpha-1)^2},
\label{eqn:aterm}\\
&&B=\frac{(\alpha-1)^2(3\Gamma-2)}{5\alpha^2-2\alpha+1}-1.
\label{eqn:bterm}
\eea
In the long wavelength limit, the spatial derivative term in 
Eq.(\ref{eqn:spert}) can be neglected and the equation is solved as
\beq
\psi=C_S({\boldmath x})+D_S({\boldmath x})\int^t_{t_0}\frac{dt}{A}.
\label{eqn:longwls}
\eeq
$C_S({\boldmath x})$ and $D_S({\boldmath x})$ are functions of spatial
coordinates.
What is expected from the analogy with the ordinary inflationary universe is 
that, as the universe expands, the decrease of $a^{-3}$ factor in $1/A$ 
keeps the integrand of Eq. (\ref{eqn:longwls}) sufficiently small so that
the long wavelength curvature perturbation stays constant.
However, this is not the case for our model.
The decrease of $\alpha$ (i.e. the conversion of the modulus kinetic 
energy to the universe expansion) overcomes the $a^{-3}$ factor, which makes
the integral term of (\ref{eqn:longwls}) no longer negligible.
This can be already seen by examining the asymptotic behaviors.
In the asymptotic past, since $A\sim t^2$, $\psi\sim 1/(-t)$ and
in the asymptotic future, $A\sim t$, hence $\psi\sim \ln t$.
In our model the variable which is constant in the long wavelength limit
in these asymptotic regions is $\phi$, the Newton potential in the uniform 
field gauge. This $\phi$, however, is not always constant but makes a
leap-like growth around the Hubble peak. 
This leap is a result of the sharp negative friction $\dot A/A$ around the 
Hubble peak. During the transition from the kinetic-driven inflation to the 
Friedmann universe, the violent movement of the background tosses up the 
perturbation of all scales.
Fig. 2 shows the behaviors of $\psi$ and $\phi$ in the long wavelength limit.

For perturbations of non-zero wavenumber $k$, the growth rate becomes more
sensitive to the background behavior. As $B$ is always negative for mild 
superinflation, the enhancement of the short wavelength perturbation is 
suppressed relatively to the long wavelength. 
For steep superinflation there appears around the Hubble peak a period where 
$B$ becomes positive, which causes an instability for short wavelength 
perturbations. The time evolutions of function $B$ for different background 
initial conditions are plotted in Fig. 3.
We characterized the background behavior by the peak value of the Hubble 
parameter, $H_{\mbox{max}}$.
While the growth of perturbation caused by $\dot A/A$ is independent of the 
wavenumber $k$, the effect of $B$ depends on the scales of the perturbations.
The {\em transfer function} of the spectrum during the graceful exit
will thus depend largely on the background behavior, i.e. the initial 
conditions. 

We have to point out the possibility
that in realistic string theory higher curvature contribution might suppress
this small scale instability.
Since this instability occurs for large $R\sim H^2$, our assumption of 
$R_{GB}^2$ dominance might not be valid. 
Thus, we cannot conclude from this analysis alone that this instability is 
inherent in the string theory.
\begin{figure}
\epsfxsize=80mm
\epsfysize=55mm
\epsffile{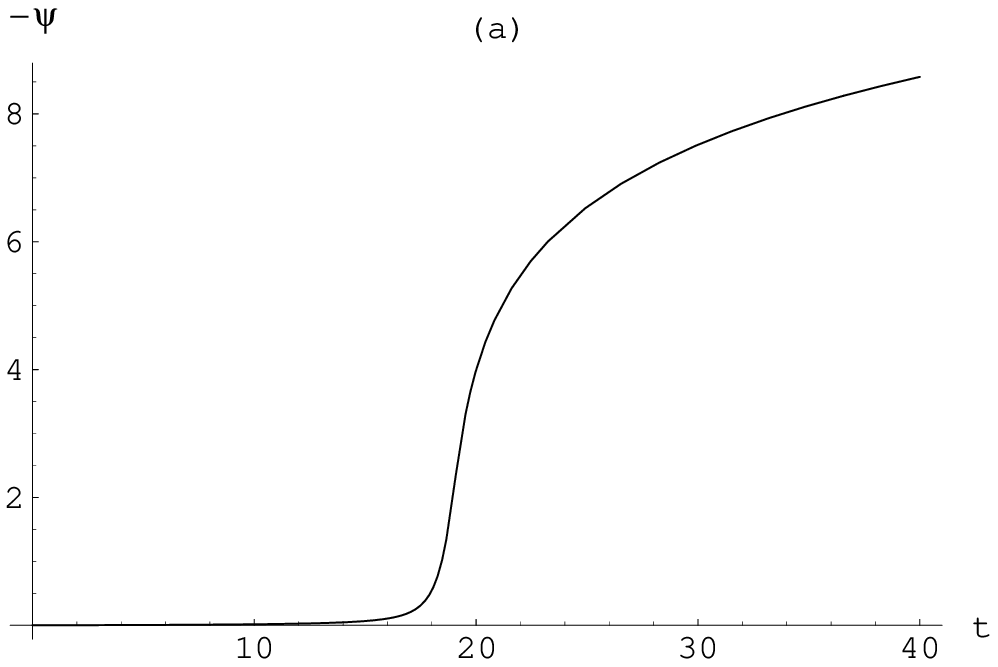}
\epsfxsize=80mm
\epsfysize=55mm
\epsffile{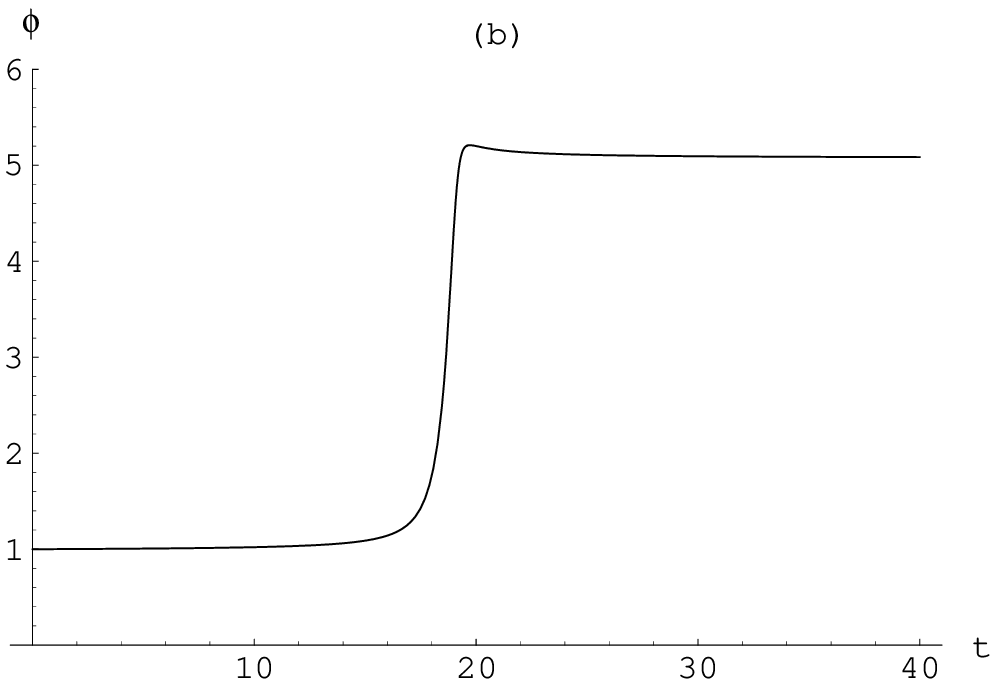}
\caption{Time evolution of long wavelength perturbations (a) $\psi$ 
and (b) $\phi$.
The background is the same as Fig. 1 ($H_{\mbox{max}}=0.5$). 
The normalization of the amplitude is set as $\phi=1$ at $t=0$. }
\end{figure}

{\em 4. Rotation and Gravitational Wave}\\
From the vector part of the perturbative equation we can see the behavior
of the vorticity. The equation can be solved to give
$S_i+a\dot F_i=0$.
This is a consequence of the absence of the rotational sources.

The tensor part of the perturbative equation is
\beq
\ddot h+\left(3H+\frac{\dot\alpha}{\alpha}\right)\dot h
+\frac{\nabla^2}{a^2}\left(5-3\Gamma\right)h=0,
\eeq
where $h$ is the Fourier component of either plus- or cross-mode gravitational 
wave amplitude.
In the long wavelength limit the equation can be solve as
\beq
h=C_T({\boldmath x})+D_T({\boldmath x})\int^t_{t_0}\frac{dt}{a^3\alpha},
\eeq
and the second term (integral) represents the growth of the perturbation
which is very similar to $\psi$ of the scalar mode. 
The short wavelength gravitational wave is quite different from the scalar
mode perturbation. Since $5-3\Gamma$ is always positive before the Hubble peak,
the gravitational wave is exponentially enhanced from the asymptotic past.
To avoid the breakdown of the background solution, initial perturbation has
to be fine-tuned to be small\cite{kss98,kss-jgrg97}.
\begin{figure}
 \vspace*{0cm}
 \hspace*{0cm}
\epsfxsize=80mm
\epsfysize=50mm
\epsffile{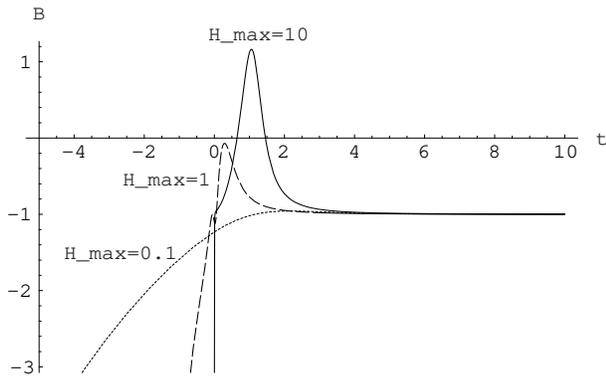}
\caption{Time evolutions of $B$ for different initial conditions. 
For mild superinflation $B$ remains negative during the transition. 
As the superinflation becomes steep, there appears a period where $B$ is 
positive, leading to an unstable mode for the short wavelength perturbations. 
The figure shows three examples, specified by the maximal values of the Hubble parameter, $H_{\mbox{max}}=0.1$, $1$, $10$ ($H_{\mbox{max}}$ is larger for 
steeper superinflation). 
For comparison $t=0$ is defined to be the time when $H=H_{\mbox{max}}$, i.e.
$\dot H=0$.}
\end{figure}


{\em 5. Discussions}\\
We have shown that during the transition there exist the enhancements of 
perturbations in a cosmological model dominated by a scalar field 
coupled to the Gauss-Bonnet term.
In the case of the dilaton-driven pre-big bang model, there is a growth
of perturbations as the universe evolves towards the curvature singularity.
While this growth is present in the longitudinal gauge, it is strongly 
suppressed if one uses the off-diagonal gauge\cite{blue}, which is supposed
to be more reliable choice of the gauge in order to avoid the breakdown
of the perturbations. 
It might be suspected that the growth of the fluctuations we found here might 
be of this kind, which can be ``gauged down.'' However, in our model the 
choice of off-diagonal gauge or longitudinal gauge will lead to more violent 
growth of the perturbation, and as can be seen from the constancy of $\phi$ 
in the asymptotic regions, the uniform field gauge we used here is supposed to
be one of the most gentle gauge choice. While the cancellation of the
growth may occur in a very accidental case, the growth we found here is a
generic result and possibility of gauging down is unlikely.
 
Although our study is limited to only one simple model, there are reasons to 
believe this is a quite general phenomena that may arise in string-cosmological
models with a successful graceful exit.
One reason is that the coupling we used here, Eq. (\ref{eqn:ssxi}), appears 
in many string low-energy effective actions as a loop correction.
Moreover, recent studies based on duality show that the function of the form
(\ref{eqn:ssxi}) appears as non-perturbative $R^2$ couplings\cite{r2coupling}, 
suggesting that this coupling form is exact beyond one-loop.
The extension of our analysis to multiple fields,
\beq
{\cal S}=\int dx^4 \sqrt{-g}
\left\{\frac 12 R-\sum_i\frac 12 \del_\mu\varphi\del^\mu\varphi_i
  -\frac{\lambda}{16}\xi(\varphi_i)R_{GB}^2
\right\},
\label{eqn:multiple}
\eeq
leads to essentially the same equations of motion for 
perturbations\cite{ssk98}.
The enhancement of the fluctuations for scalar and tensor modes is a 
consequence of non-negligible $1/a^3\alpha$, which means the rapid conversion 
of the scalar field kinetic energy into the kinetic energy of the universe
expansion. Since this situation is the same for any cosmological model
with a transition from a superinflation to Friedmann universe, we expect 
the existence of the same enhancement of the perturbations even for more 
realistic models with dilaton or other fields, as long as the graceful exit 
is accomplished through the same mechanism. The model-independent approach
by Brustein et.al.\cite{brustein} is of interest from this viewpoint.

We are grateful to Jai-chan Hwang and Hyerin Noh for the discussions in 
19th Texas Symposium in Paris. We also appreciate the useful discussions with 
Atsushi Taruya.
Part of this work by J.S. is supported by the Grant-in-Aid for 
Scientific Research No. 10740118.



\end{document}